\documentclass{article}

 \usepackage{arxiv}
 
 \usepackage[utf8]{inputenc} 
 \usepackage[T1]{fontenc}    
 \usepackage{hyperref}       
 \usepackage{url}            
 \usepackage{booktabs}       
 \usepackage{amsfonts}       
 \usepackage{nicefrac}       
 \usepackage{microtype}      
 \usepackage{lipsum}     
 \usepackage{graphicx}
 \usepackage[numbers]{natbib}
 \usepackage{doi}

\title{Multi-scale simulation of thrombus formation at LVAD inlet cannula connection: Importance of Virchow’s triad}

\author{
  Rodrigo M\'endez Rojano \\
  Meinig School of Biomedical Engineering \\
  Cornell University\\
  Ithaca, NY 14850 \\
  \texttt{rm2235@cornell.edu} \\
   \And
  Mansur Zhussupbekov \\
  Meinig School of Biomedical Engineering \\
  Cornell University\\
  Ithaca, NY 14850 \\
  \texttt{mz332@cornell.edu} \\
  \And 
  James F. Antaki\\
  Meinig School of Biomedical Engineering \\
  Cornell University\\
  Ithaca, NY 14850 \\
  \texttt{antaki@cornell.edu} \\
}

\begin{document}
\maketitle

\begin{abstract}
As pump thrombosis is reduced in current-generation ventricular assist devices (VAD), adverse events such as bleeding or stroke remain at unacceptable rates. Thrombosis around the VAD inlet cannula (IC) has been highlighted as a possible source of stroke events. Recent computational fluid dynamics (CFD) studies have attempted to characterize the thrombosis risk of different IC-ventricle configurations. However, purely CFD simulations relate thrombosis risk to ad-hoc criteria based on flow characteristics, with little consideration of biochemical factors. This study investigates the genesis of IC thrombosis including two elements of the Virchow’s triad: Endothelial injury and Hypercoagulability. To this end a multi-scale thrombosis simulation that includes platelet activity and coagulation reactions was performed.  Our results show significant thrombin formation in stagnation regions ($|\textbf{u}|< 0.002$ m s$^{-1}$) close to the IC wall. In addition, high shear-mediated platelet activation was observed over the leading-edge tip of the cannula which mirrors the thrombus deposition pattern observed clinically. The current study reveals the importance of biochemical factors to the genesis of thrombosis at the ventricular-cannula junction which can inform clinical decisions in terms of anticoagulation/antiplatelet therapy and guide engineers to develop more robust designs. 
\end{abstract}

\keywords{Multi-Scale Thrombosis Modeling, Ventricular Assist Devices, Inlet Cannula, tissue factor, coagulation }

\section{Introduction} \label{sec:Introduction}
Current-generation ventricular assist devices (VADS) have virtually eliminated the incidence of pump thrombosis \cite{Mehra2019}. However, adverse events related to coagulation such as bleeding or stroke remain at unacceptable rates \cite{Colombo2019,Krabatsch2017,McIlvennan2019}.\\ 
Thrombosis around the VAD inlet cannula (IC) has been recently highlighted as a possible source of ischemic stroke events in patients receiving a Left Ventricular Assist Device (LVAD) \cite{Glass2019,Ghodrati2020}. In fact, the potential role of the ventricular cannula in thrombosis related complications has motivated numerical, in-vitro, and clinical studies for over two decades \cite{Ghodrati2020a,Antaki1995,Curtis1998,Bachman2011,Bhama2009,Drummond2009,Sorensen2013,Rossini2020}.\\  

The majority of studies have concentrated in the intraventricular flow dynamics, specifically in the region of the ventricular apex. Early flow visualization studies in-vitro \cite{Antaki1995,Curtis1998} and in isolated hearts revealed the existence of stagnation and recirculation regions related to the shape and positioning of the cannula tip \cite{Sorensen2013}. More recently, particle image velocimetry studies were conducted on the EVAHEART (EVAHEART, Inc., Houston, TX) cannula by May-Newman et al. \cite{May-Newman2019,May-Newman2019a} and postulated that systematic washout may reduce the risk of thrombosis.\\  

Chivukula et al. \cite{Chivukula2018,Chivukula2018a,Chivukula2019a} studied the thrombosis risk in IC-LV configurations using a computational fluid dynamics (CFD) approach. Using a Lagrangian approach, Chivukula and coworkers assessed the influence of the cannula angle, the cannula protrusion height, and the ventricle size on the thrombosis risk. In a similar vein, Liao et al. \cite{Liao2016} studied the influence of cannula geometry on blood washout and hemolysis index. Recently, Ghodrati et al. \cite{Ghodrati2020a,Ghodrati2020} performed transient simulations in patient-specific, cannulated left ventricles to investigate the influence of the insertion site position and orientation, with the goal of relating stagnation regions and non-physiological wall shear stress to thrombosis risk. Most recently, Rossini et al., employed 2D color doppler velocimetry in seven LVAD patients with dilated cardiomyopathy to investigate anomalies in intraventricular flow that may increase the risk of pump-related thrombosis \cite{Rossini2020}.\\ 

Although the aforementioned numerical and experimental studies have provided great insight into the IC thrombogenesis risk, they all have relied upon ad-hoc, qualitative criteria to assess thrombosis risk. None, to the best knowledge of the authors, have considered the biochemical elements that contribute to the formation and stabilization of the thrombus.\\   

This study aimed to investigate the genesis of IC thrombosis considering Virchow’s triad, employing a multi-scale thrombosis simulation that includes cellular and biochemical transport, as well as coagulation reactions. Our approach considered biochemical transport, platelet deposition, cleaning, and activation as well as coagulation reactions triggered by tissue factor. The aim of the current work is to provide a better insight into the mechanisms that drive thrombosis at the IC and identify the biochemical mechanisms that promote thrombus growth.\\

\section{Methods}
\label{sec:Methods}

A set of chemo-fluidic multi-scale CFD simulations using the model of Wu et al. \cite{Wu2017} were performed considering a VAD IC within an anatomically correct dilated left ventricle (LV). Details of the ventricular geometry have been described previously in \cite{TimothyBachman2018}.  

\subsection{Thrombosis Model}

\begin{figure}[h!]
\centering
\includegraphics[width=0.9\textwidth]{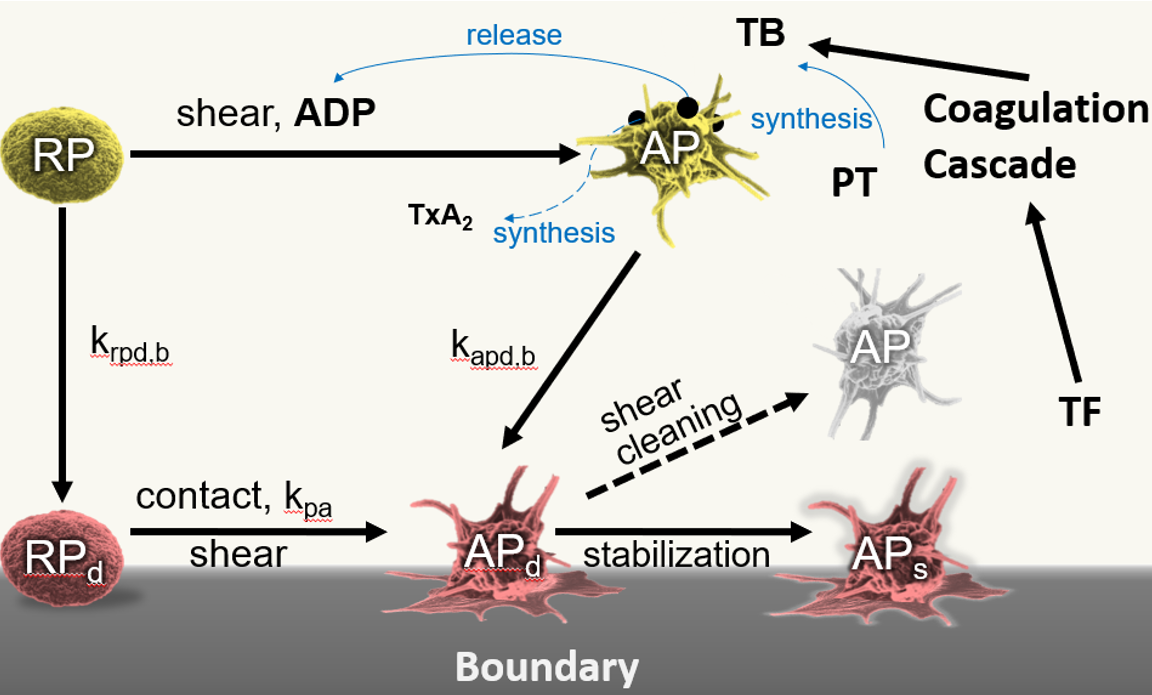}
\caption{Thrombosis model diagram describing the platelet-based model of Wu et al. \cite{Wu2017}. In addition, TF-driven thrombin (TB) production is depicted.\label{fig:1} }
\end{figure}

The thrombosis model of Wu et al. \cite{Wu2017} was used for the computations. In this framework, the Navier-Stokes (NS) equations for blood and scalar equations for its biochemical constituents are solved in a coupled manner. Mechanical and chemical activation of platelets can be induced by shear stress or the combined effect of adenosine diphosphate (ADP), thrombin (TB), and thromboxane A2 (TxA2). Platelets can deposit onto artificial surfaces or on previously deposited platelets forming a thrombus. Deposited platelets and thrombus periphery are susceptible to cleaning by local wall shear stress. The growing thrombus can alter the flow field by means of a hindrance term in the momentum equation in regions where the platelet volume fraction is high. (See Fig.~\ref{fig:1}) For a more detailed description of the present thrombosis model and thrombosis modeling in general, the reader is referred to \cite{Wu2017,Fogelson2015}. 

\subsection{Baseline Geometry and boundary conditions}
\begin{figure}[h!]
\centering
\includegraphics[width=0.9\textwidth]{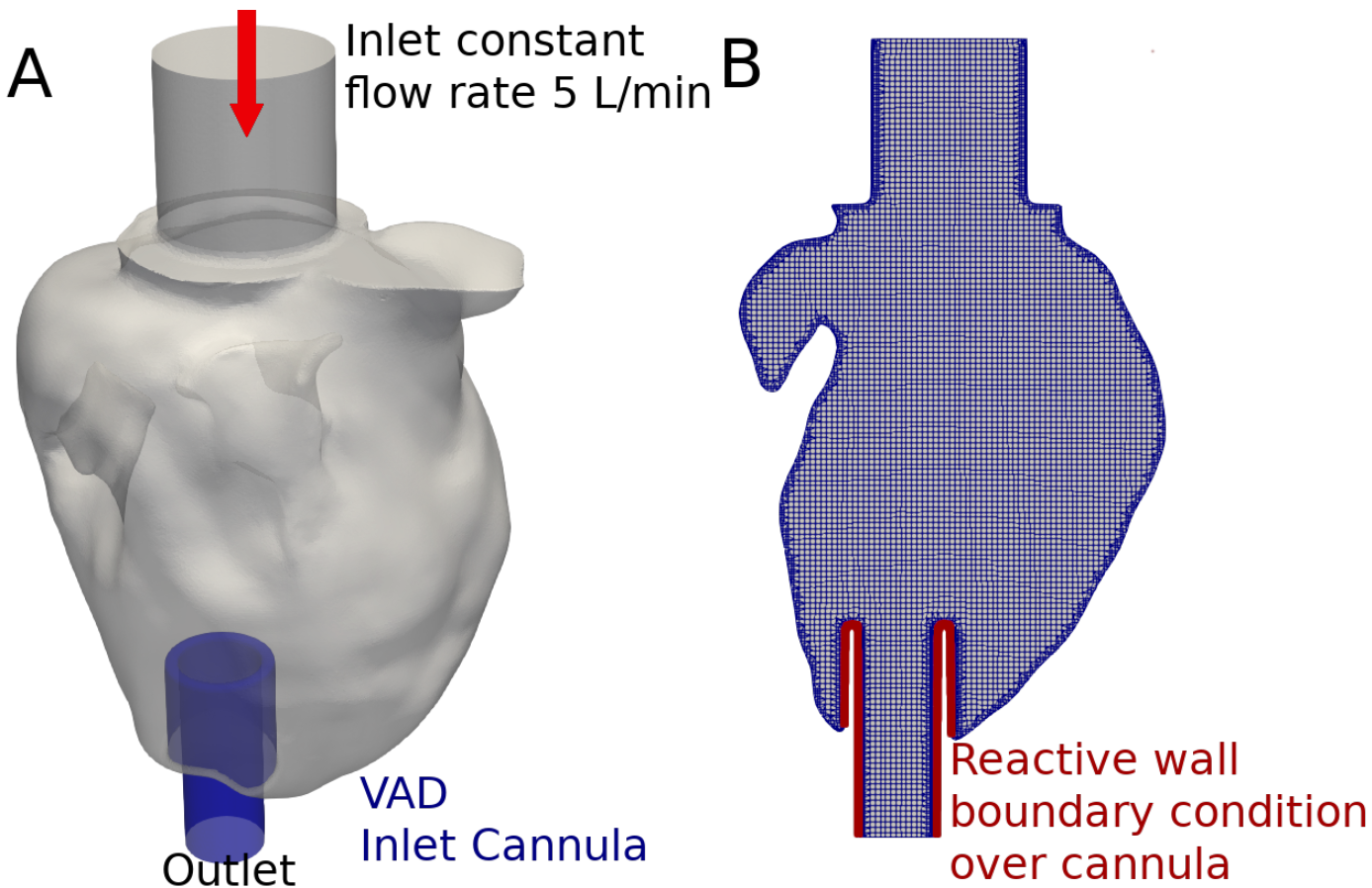}
\caption{A) Computational domain showing the dilated LV and the LVAD inflow cannula inserted at the apical position. B) Middle cut visualization of the unstructured mesh visualization used in the thrombosis simulations. \label{fig:2}}
\end{figure}

Figure~\ref{fig:2}A shows the computational domain and mesh used in the simulations. A generic VAD cannula tip was placed in-silico at the apex, along the long axis of an anatomically correct dilated human LV. The outer and inner diameter of the IC are 20 mm and 14 mm, respectively. The insertion depth of the IC in the LV was about 20 mm. At the atrium, a conduit of 30 mm diameter and 30 mm height was added to promote a developed velocity profile and represent the out-flow jet typical to this configuration \cite{Ghodrati2020}. A uniform velocity boundary condition was prescribed at the inlet of the entrance region (corresponding to constant flow rate of 5 L min$^{-1}$). Non-slip boundary condition was set at the walls of the LV and the inflow cannula. For the biological scalar species, physiological concentration of platelets was prescribed $3 \times 105$ platelets $\mu$L$^{-1}$ at the inlet with 1\% of background platelet activation. A reactive boundary condition was used at the inflow cannula surface allowing both platelet deposition and cleaning on the walls.\\

Figure~\ref{fig:2}B shows the corresponding unstructured mesh with 840k hexahedral and tetrahedral elements that was used in the simulations. The time step used in the simulations was 0.0001 s which ensured that CFL $<$ 1. A mesh convergence study was performed using progressively finer meshes of 500k, 840k, and 2.5 million elements, which indicated that the computed IC wall shear stress was sufficiently accurate with 840k elements. The fluid was assumed Newtonian with a kinematic viscosity of $3.333 \times 10^{-6}$ m$^2$ s$^{-1}$ and a density of 1050 kg m$^{-3}$. The solution of the flow and scalar equations was performed using the open-source CFD software OpenFOAM~ \url{www.openfoam.org}.\\   

The species considered in the thrombosis model are the following: activated platelets (AP), resting platelets (RP), prothrombin (PT), thrombin (TB), antithrombin (AT), thromboxane A2 (TxA2), and adenosine diphosphate (ADP). A specific reaction term is solved for each of the species according to their biochemical interactions. For an extensive description of the reaction terms and diffusion coefficients of the biochemical species, see \cite{Wu2017}. The initial conditions and specific model parameters used in the present study are listed in Table~\ref{tab:1}.\\

\begin{table}[h!]
\centering
\begin{tabular}{|c|c|}
\hline
\textbf{Parameter}&\textbf{Value}\\
\hline
\hline
Inlet AP concentration [Plt m$^{-3}$] & $3.0 \times 10^{12}$ \\
Inlet RP concentration [Plt m$^{-3}$] & $3.0 \times 10^{14}$ \\
PT concentration [M] & $1.1 \times 10^{6}$ \\
ATIII concentration [M] & $2.844 \times 10^6 $ \\
RP-boundary(wall) deposition rate, karpd [m s$^{-1}$] & $4.0 \times 10^{-5}$ \\
AP-boundary(wall) deposition rate, kaapd [m s$^{-1}$] & $4.0 \times 10^{-4}$ \\
\hline
\end{tabular}
\caption{Initial conditions and thrombosis model parameters. See Wu et al. \cite{Wu2017} for definitions of these terms and their use in the species convection-diffusion-reaction equations.\label{tab:1}}
\end{table}
 
\subsection{Coagulation reactions triggered by Tissue Factor}

\begin{figure}[h!]
\centering
\includegraphics[width=0.6\textwidth]{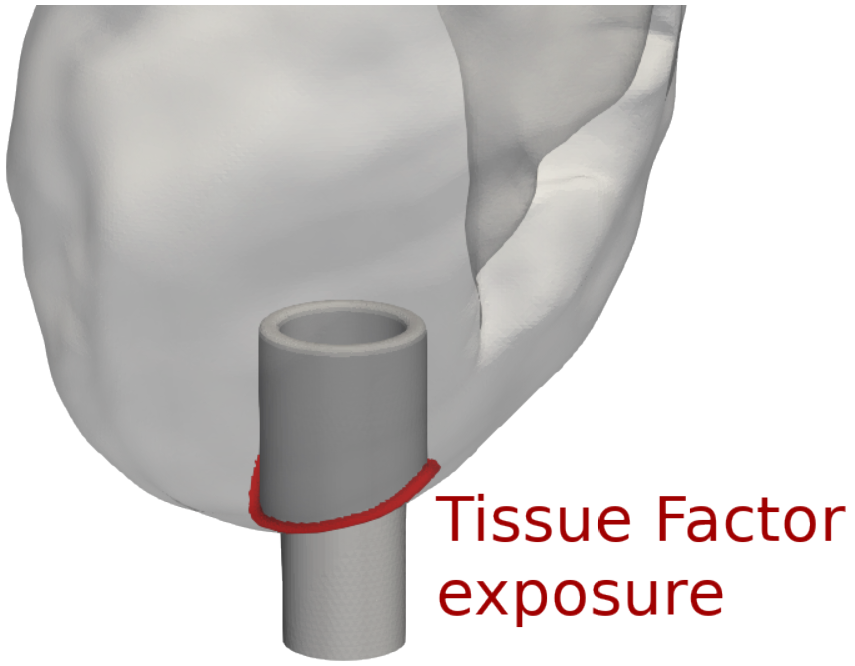}
\caption{TF exposure boundary condition region (red) at the junction of the cannula and the myocardial surface of dilated LV.  \label{fig:3}}
\end{figure}

During VAD implantation, a coring procedure is performed at the apex of the LV to insert the IC. This procedure can potentially result in exposed myocardium within the ventricle at the interface with the cannula tip. Consequently, tissue factor (TF) is exposed, which may initiate a cascade of enzymatic reactions that leads to thrombin production which participate in various feedback pathways that contribute to the formation and stabilization of thrombus. Accordingly, an “injury” region adjacent to the ventricle-cannula interface was introduced to the domain (See Fig.~\ref{fig:3}). Over this region, a Dirichlet boundary condition with a constant concentration value of TF = 0.8 nmol m$^{-3}$ was prescribed.  

\begin{figure}[h!]
\centering
\includegraphics[width=0.5\textwidth]{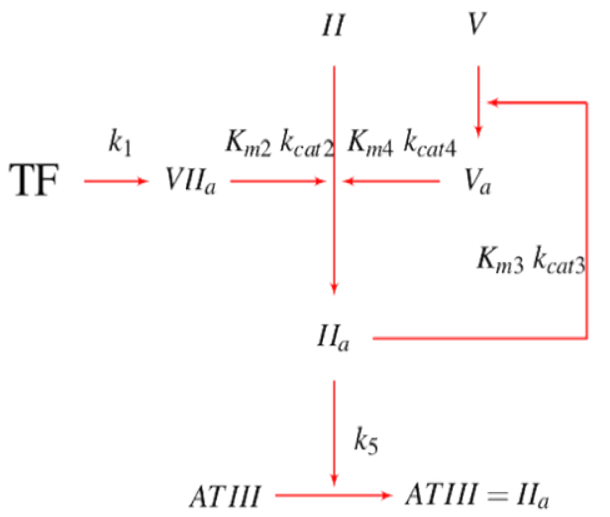}
\caption{Reduced biochemical scheme for thrombin production triggered by Tissue Factor (TF) based on \cite{MendezRojano2019}. Factor VIIa activates prothrombin (II), then thrombin (IIa) activates factor V in a feedback loop and is consumed by antithrombin (ATIII). \label{fig:4}}
\end{figure}

To model the dynamics of thrombin formation, a simplified set of five additional reactions were incorporated into the existing thrombosis model (See Fig.~\ref{fig:4}). This reduced-order scheme economizes computational time by using a minimal number of biochemical species \cite{MendezRojano2019}.

\begin{figure}[h!]
\centering
\includegraphics[width=0.7\textwidth]{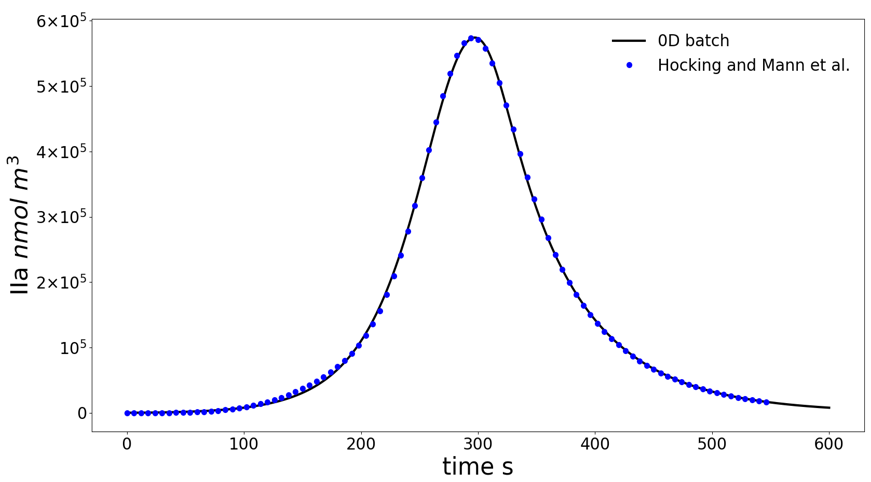}
\caption{Thrombin production in batch reactor simulation. Current model results are compared with the results published by Hockin et al.\cite{Hockin2002}.\label{fig:5}}
\end{figure}

This approach has been used in literature to study the biochemistry of coagulation along with complex fluid dynamics. For the current application, the hallmark model of Hockin et al. \cite{Hockin2002} was used as a reference to calibrate the reduced model. Least square optimization was performed with the Python module S-timator (\url{https://webpages.ciencias.ulisboa.pt/~aeferreira/stimator/}) to identify parameters that best fit the numerical data of Hockin et al. (See Table~\ref{tab:2}). Figure~\ref{fig:5} shows the thrombin formation curve for both the reduced kinetic model and the data from \cite{Hockin2002}. \\

\begin{table}[h!]
\centering
\begin{tabular}{|c|c|}
\hline
Parameter & value\\
\hline
\hline
k1 & $ 2.28378 \times 10^{-5}$ s$^{-1} $ \\
km2 & $30.1908 \times 10^3$ nmol m$^{-3}$ \\
kcat2 & 1.09473 s$^{-1}$ \\
km3 & $7.90731 \times 10^3$ nmol m$^{-3}$ \\
kcat3 & $0.00178566$ s$^{-1}$ \\
km4 & $202.974 \times 10^3$ nmol m$^{-3}$ \\
kcat4 & $1.17018$ s$^{-1}$ \\
kcat5 & $7.24351 \times 10^{-9}$ nmol m$^{-3}$ s$^{-1}$ \\
II$_0$ & $1610.98 \times 10^3$ nmol m$^{-3}$ \\
AT$_0$ & $3576.03 \times 10^3$ nmol m$^{-3} $ \\
V$_0$ & $20 \times 10^3$ nmol m$^{-3} $ \\
\hline
\end{tabular}
\caption{Kinetic parameters and coagulation factors and initial conditions. \label{tab:2}}
\end{table}

Excess of calcium and phospholipids are assumed for both the batch reaction and free flow cases. The thrombin produced by the coagulation reactions was coupled with the Wu platelet-based model by means of thrombin-mediated platelet activation. Thus, this new source of thrombin contributes to platelet activation in regions where the thrombin activation threshold is reached. 
 
\section{Results}
\label{sec:results}
\subsection{Intraventricular Flow Field}
\begin{figure}[h!]
\centering
\includegraphics[width=0.7\textwidth]{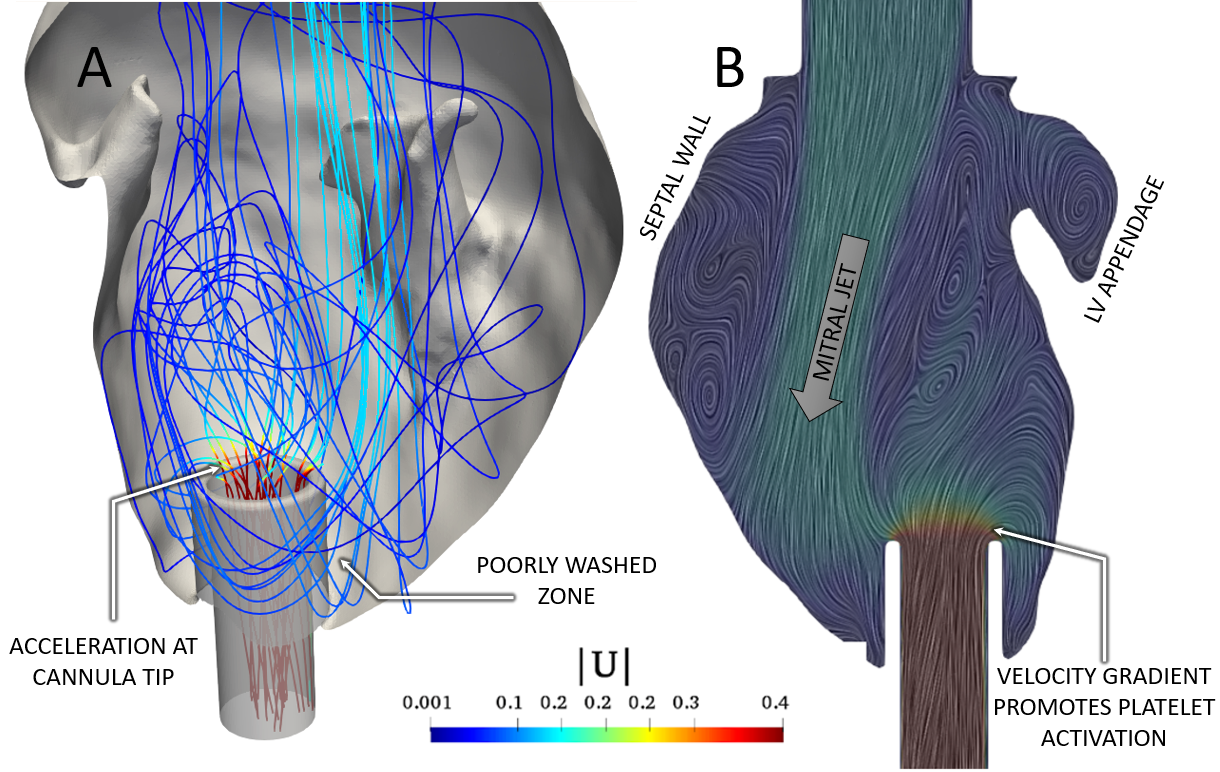}
\caption{A) Path lines inside the dilated LV B) Middle cut velocity magnitude field.\label{fig:6}}
\end{figure}

Figure~\ref{fig:6} A shows the streamlines inside the LV. Relatively low flow velocity regions can be observed around the IC. Streamlines near the cannula leading edge show important velocity gradients which results in elevated mechanical platelet activation.  Figure~\ref{fig:6} B shows the velocity field magnitude in the coronal plane, highlighting the presence of the mitral jet. Due to the ideal position of the cannula (parallel to the septal wall), the mitral jet plays an important role in washout of platelet agonists near the cannula. In general terms, the velocity field compares well qualitatively with previous numerical and experimental studies \cite{Curtis1998,May-Newman2019,Ghodrati2020}. Two regions of low shear stress (< 0.005 Pa) appeared at the root of the IC (see Fig.~\ref{fig:7} A). Biochemical species advected or diffused inside these regions (stasis regions) may promote platelet activation, further platelet deposition and hence thrombus formation. In regions where sufficient blood wash out is achieved the concentration of agonist is lower limiting further thrombus growth.\\

\subsection{Platelet activity}
\begin{figure}[h!]
\centering
\includegraphics[width=0.8\textwidth]{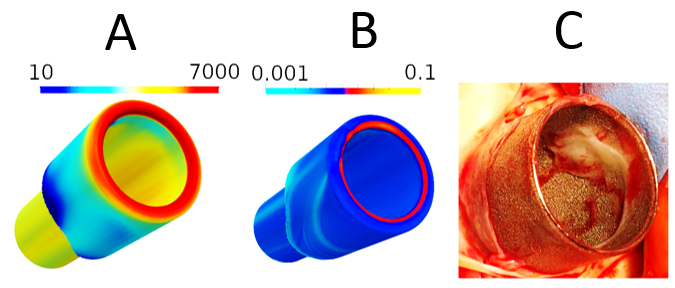}
\caption{LV IC A) shear rate field (s$^{-1}$), B) Volume fraction of deposited platelets (at 150s), HM3 clinical explant (courtesy of Dr. B. P. Griffith).\label{fig:7}}
\end{figure}

The shear stress magnitude within most of the ventricle was less than 10 Pa which did not provoke platelet activation. However, at the inner leading edge of the cannula tip, shear rate of up to 7000 s$^{-1}$ was observed, resulting in mechanical shear activation (See in Fig.~\ref{fig:7} A). For the first 150 seconds, platelets accumulated along the radial rim of the IC as shown in Fig.~\ref{fig:7} B. However, following 150 seconds, platelet volume fraction reached a plateau, due the equilibrium between deposition and shear cleaning (washing). These simulated results resembled the thrombus deposition pattern observed clinically from a cannula explanted from a Heatmate-3 patient (See Fig.~\ref{fig:7} C).\\

\begin{figure}[h!]
\centering
\includegraphics[width=0.6\textwidth]{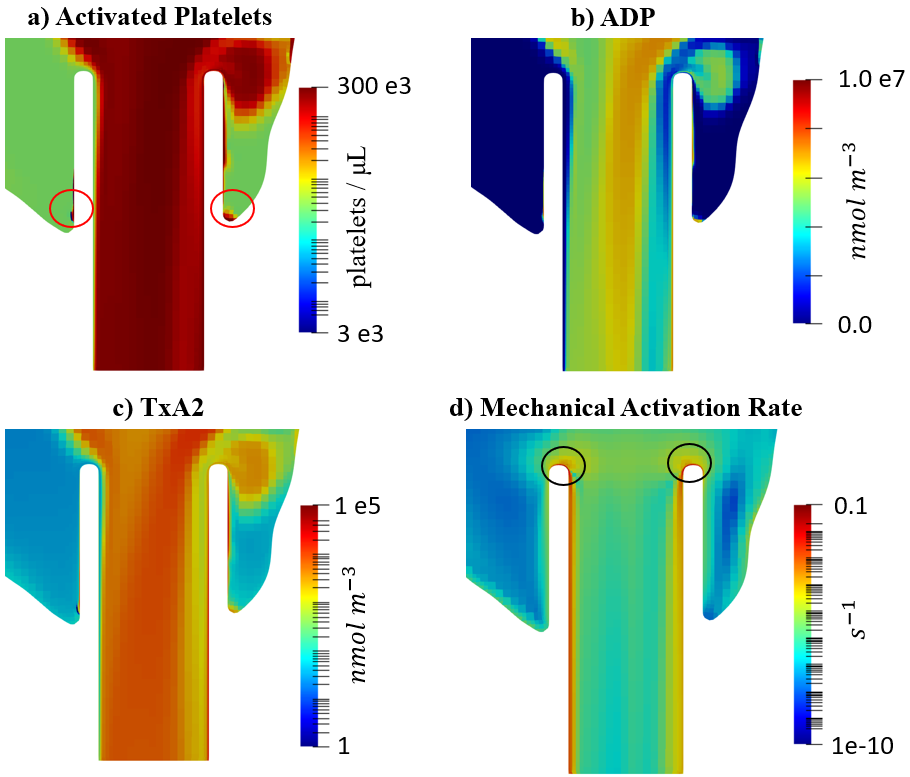}
\caption{Coronal plane visualization of key scalar fields in simulation at 150s, (a) activated platelets, (b) ADP, (c) Thromboxane, and (d) mechanical platelet activation rate. Activated platelets at cannula-ventricular junction (red circles) lead to aggregation. Mechanical activation initiated at leading edge of cannula tip (black circles). \label{fig:8}}
\end{figure}

Most activated platelets and the agonists released in the process (ADP and TxA2) are aspirated inside the VAD (see Fig.~\ref{fig:8} A –~\ref{fig:8}C). Few activated platelets enter a stasis region at the base of the cannula near the junction with the endocardium. The middle coronal plane visualization reveals increased mechanical platelet activation at the cannula leading edge and the inner cannula surface.\\

\subsection{Thrombin production}

\begin{figure}[h!]
\centering
\includegraphics[width=0.6\textwidth]{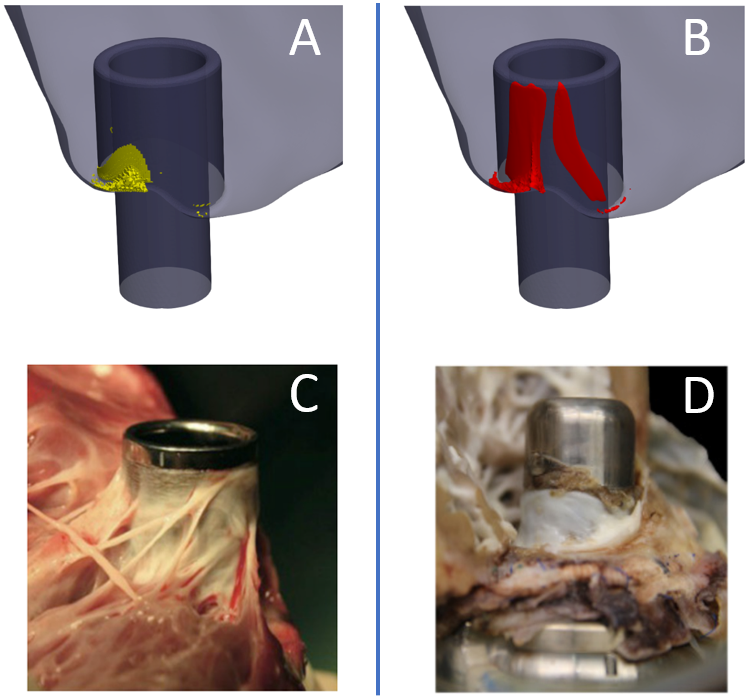}
\caption{A) Velocity magnitude threshold $|\textbf{u}| < 0.002$ m s$^{-1}$. B) Thrombin production due to tissue factor exposure; concentration threshold thrombin > 1 nmol m-3 at 150 s. C) Neo intima tissue growth around EVAHART cannula \cite{Yamada2011a}. D) Thrombus growth in HVAD IC \cite{Glass2019}.\label{fig:9}}
\end{figure}

Simulated thrombin synthesis due to coagulation reactions and activated platelets resulted in super-critical thrombin concentration in the low-velocity regions ($|\textbf{u}| < 0.002$ m s$^{-1}$) as shown in Fig.~\ref{fig:9} A and B. Figure~\ref{fig:9} B shows the regions where thrombin concentration exceeds 1 nmol m$^{-3}$. This means that the characteristic flow residence time was large enough for coagulation reactions to take place and promote prothrombin transformation. As expected, the regions in which significant concentration of thrombin occurs coincides with low wall shear stress regions that have poor flow wash out. The origin thrombin produced in these regions was coagulation reactions rather than upstream platelet activation by mechanical shear. The simulated thrombin concentration pattern was found to resemble neo intimal tissue from explanted cannula of the EVAHEART Fig.~\ref{fig:9} C and thrombus deposited on the HVAD cannula Fig.~\ref{fig:9} D. This could be explained by the critical role of thrombin in the cleavage of fibrin. 

\begin{figure}[h!]
\centering
\includegraphics[width=0.8\textwidth]{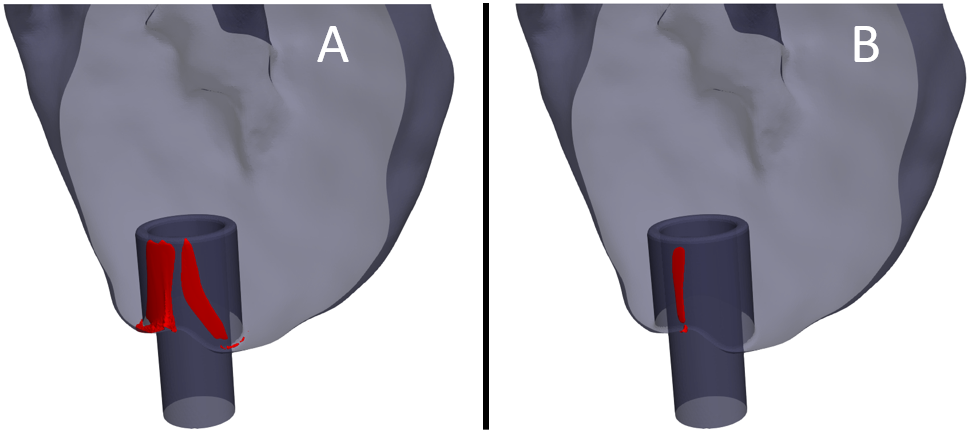}
\caption{A) Thrombin threshold region (1 nmol m${-3}$) at 150 seconds using a constant TF Dirichlet boundary condition. B) Thrombin threshold region (1 nmol m$^{-3}$) at 150 seconds Dirichlet TF boundary condition turned off after 50 seconds. \label{fig:10}}
\end{figure}

Figure~\ref{fig:10} A illustrates the region adjacent to the cannula tip corresponding to thrombin concentration greater than 1 nmol m$^{-3}$. To investigate if the production of thrombin would continue following a brief exposure of TF, an additional simulation was performed in which the TF boundary condition was set to zero after 50 s. In this case, thrombin production was substantially reduced and the accumulated thrombin and enzymes near the cannula tip region were rapidly cleared (See Fig.~\ref{fig:10} B). 

\section{Discussion}\label{sec:discussion}

Numerical simulation of thrombosis has become a viable tool to predict thrombus formation in blood wetted medical devices.  Advances in VAD technology have reduced the incidence of pump thrombosis; nonetheless, ischemic stroke remains a significant adverse event in VAD patients. Recent studies have pointed to the IC as a possible source of emboli and subsequent stroke \cite{Ghodrati2020,Glass2019}. CFD studies of left ventricles and IC have identified stagnation regions and flow patterns which are related to an elevated thrombosis risk. However, most CFD analyses rely on qualitative assessment of the velocity and shear fields, which are somewhat arbitrary or ad-hoc. To the best knowledge of the authors, no prior study has explicitly simulated the genesis of thrombus, which is a complex process involving biochemical reactions and transport of platelets and agonists, particularly thrombin. In this study, the three components of the Virchow’s triad were considered: flow, endothelial injury, and coagulation dynamics.  In terms of platelet activation, our results suggest that non-physiological shear stress rates present at the inner radius of the cannula activates platelets that deposit downstream on the interior surface, resembling thrombosis patterns found on cannulas of explanted VADs.\\ 

At the junction between the LV apex and the IC, TF-driven thrombin formation was sustained until the end of the simulation (150 s). Thrombin accumulated mainly in low-velocity regions ($|\textbf{u}| < 0.002$ m s$^{-1}$), which is concordant with criteria used in previous CFD studies. For example, Ghodrati et al. (2020) specified a velocity magnitude of $0.005$ m s$^{-1}$ as a theshold for thrombosis risk \cite{Ghodrati2020}.  Thrombin formed in these regions can lead to fibrin polymerization and activation of platelets, which further amplifies thrombin generation. Agonists accumulated via this positive feedback loop may be transported up and over the cannula wall, whereupon pre-sensitized platelets (due to high shear) may become activated and deposit downstream. As more attention is given to prevention of blood stasis regions during LVAD support \cite{Sorensen2013}, it is imperative to build stasis criteria upon biochemical mechanisms that are less arbitrary than a prescribed cut-off value. A brief exposure of TF was explored in the simulations showing that thrombin formation can be limited by shutting down the influx of TF. This phenomenon suggest that the thrombin formation process can be altered by suppressing the initial step of the coagulation cascade, which might be an interesting anticoagulation target.  Platelet deposition in low wall shear stress zones near the base of the cannula continued to grow slowly over the course of the simulation. However, the rate of growth was limited by the reduced platelet transport towards this region. In the long run, the accumulation of pro-coagulant species such as activated platelets or thrombin towards stagnation regions could be a potential source of thrombus at the base of the cannula. However, the computational cost of this simulation prevented us from reaching that timepoint.\\

Despite the insightful results of these simulations, we acknowledge several limitations of the current study. First, the surface properties (material composition, chemistry, texture) of the cannula itself was not explicitly prescribed. Instead, this study focused on the effect of exposed tissue factor at the IC insertion site due to surgical wound. Another limitation was the absence of pulsatility and LV wall motion, both of which may play a role in the dynamics of species transport and thrombus washout. However, the combined computational cost of fluid-structure interaction, pulsatility (unsteady flow), and the multi-constituent thrombosis model would be prohibitive. Nevertheless, simulations on a static ventricle reflect a worst-case scenario, corresponding to a severe dilated cardiomyopathy \cite{Chivukula2018,Chivukula2018a,Chivukula2019a,Liao2016}. These simulations also did not account for fibrin polymerization which might be important in thrombus stabilization/embolization mechanisms. This is a topic of an ongoing investigation. Lastly, the current study considers the IC positioned along the long axis of the ventricle; however, cannula misalignment is common in LVAD patients, which could exacerbate the thrombogenic processes described here. Extension of these studies to account for sub-optimal cannula positioning and various anticoagulation strategies is a focus of ongoing work.\\  

In conclusion, the genesis of thrombus deposition at the ventricular-cannula junction relies on a complex interaction of hemodynamic, hematological, and biochemical factors – elements of Virchow’s triad. Accordingly, this study aimed to elucidate the importance of introducing the synthesis and transport of biochemical species into a simulation in combination with intraventricular fluid dynamics.  We hope that continued investigation along these lines will contribute to the utility of in-silico studies in guiding the design of ventricular apical cannula and postoperative anticoagulation management, which can help reduce the incidence of embolization and attendant adverse events in VAD patients.\\  

\section{Acknowledgments} \label{sec:acknow}
This research was supported by NIH R01 HL089456 and NIH R01 HL086918.

\section{Disclosure}\label{sec:disclosure} 
The authors declare no conflict of interest. 

\bibliographystyle{unsrt}  
\bibliography{ref.bib}  

%
%
%
%

\end{document}